\documentstyle[11pt,newpasp,twoside]{article}
\def\edcomment#1{\iffalse\marginpar{\raggedright\sl#1\/}\else\relax\fi}
\reversemarginpar

\begin{document}
\title{Synthetic Color-Magnitude Diagrams for $\omega$~Centauri and Other
Massive Globular Clusters with Multiple Populations}
\author{Chang H. Ree, Suk-Jin Yoon, Soo-Chang Rey, and Young-Wook Lee}
\affil{Center for Space Astrophysics, Yonsei University, Seoul 120-749, Korea}

\begin{abstract}
We have constructed synthetic color-magnitude diagrams (CMDs) for $\omega$~Cen
and other massive globular clusters with apparently peculiar CMD morphology.
Our population models, which adopt the most up-to-date input physics and
parameters, show that the observed CMD of $\omega$~Cen can be reproduced by
adopting (1) multimodal metallicity distribution function as derived from
the observed color distribution of red-giant-branch (RGB) stars, and (2) an
internal age-metallicty relation among the populations therein.
Similar results were obtained for other massive globular clusters with bimodal
horizontal-branches (HBs). In particular, we found that the peculiar CMD
morphology (broad RGB, bimodal HB) and properties of RR Lyrae stars observed
in NGC~6388 and NGC~6441 can be reproduced by the composite of two distinct
populations with mild internal age-metallicity relations. This suggests that
these clusters, as well as $\omega$~Cen, may represent the relicts of disrupted
dwarf galaxies.
\end{abstract}

\section{Introduction}
Our recent study on the most massive globular cluster (GC) $\omega$~Cen revealed
the presence of several distinct stellar populations and internal age-metallicty
relation among them (Lee et al. 1999).
This result, together with the fact that the 2nd most massive GC, M54, is in
fact the nucleus of the Sagittarius dwarf galaxy (Layden \& Sarajedini 2000),
has motivated us to investigate other massive Galactic GCs with peculiar
color-magnitude diagram (CMD) morphologies to see whether they also have any
evidence for multiple stellar populations.
In this paper, we report our progress in the detailed modeling of $\omega$~Cen
and two other massive GCs, NGC~6388 and NGC~6441.

\section{$\omega$~Centauri}

\begin{figure*}
\includegraphics{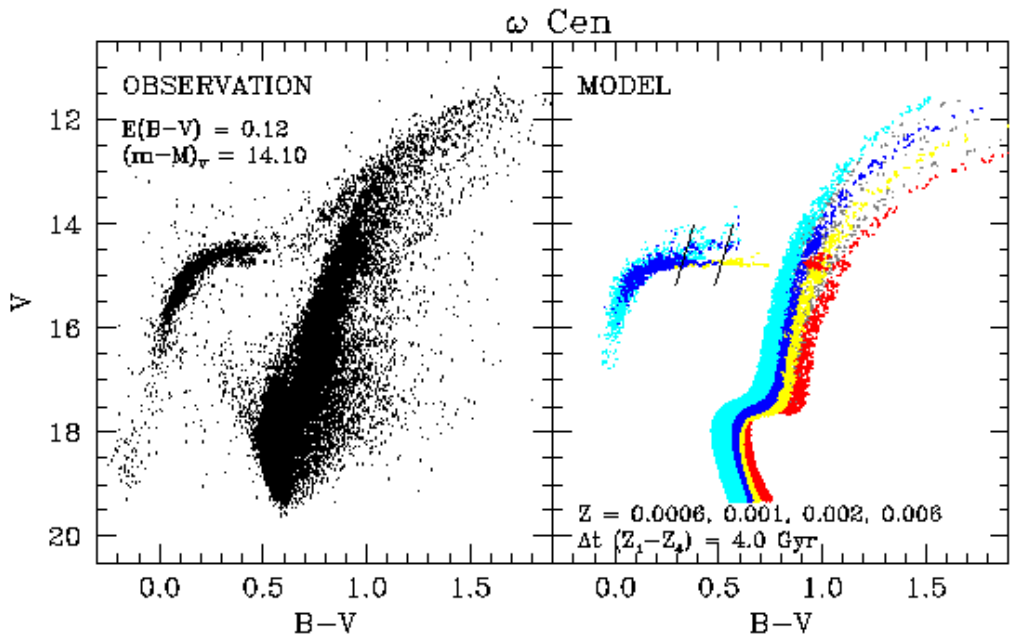}
\vspace{9cm}
\caption{$Left$: Observed CMD of $\omega$~Cen. $Right$: Synthetic CMD for 
$\omega$~Cen with four distinct stellar populations.}
\end{figure*}

Figure 1 compares the observed CMD (Lee et al. 1999; Rey et al., this volume)
and our best population model for $\omega$~Cen.
The models were constructed with the most recent $Y^2$ isochrones and HB tracks
(Yi et al. 2001).
The metallicities of 4 major components were chosen from the metallicity distribution
function (MDF) of bright red-giant-branch (RGB) stars (Rey et al., this volume).
The age difference between the oldest (and the most metal-poor) and the youngest
(and the most metal-rich) populations is predicted to be $\sim$ 4 Gyr.
Such an age difference comes from the age dating technique utilizing 
horizontal-branch (HB) morphology difference between the most metal-rich
and the most metal-poor populations.
An illustration of our age dating technique is shown in Figure 2.
With fixed metallicites found in the MDF, the HB morphologies (HB mean colors)
and the main-sequence (MS) turnoff luminosities vary with the relative ages 
between the two populations.
The red clump associated with the most metal-rich population is partially
overlapped by the metal-poor RGB bump stars (see Rey et al., this volume), and
our models best reproduce the location of the red clump when the age of the
most metal-rich population is some 4 Gyr younger than the most metal-poor
population.
Thus the stellar populations in our model for $\omega$~Cen follow an
age-metallicity relation spanning from $Z$ $\sim$ 0.0006 to 0.006 in metallicity,
and $\Delta$t $\sim$ 4 Gyr in age.
Age differences greater than 6 Gyr are also excluded from the turnoff morphology
(i.e., metal-rich turnoff becomes too bright).

\begin{figure*}
\includegraphics{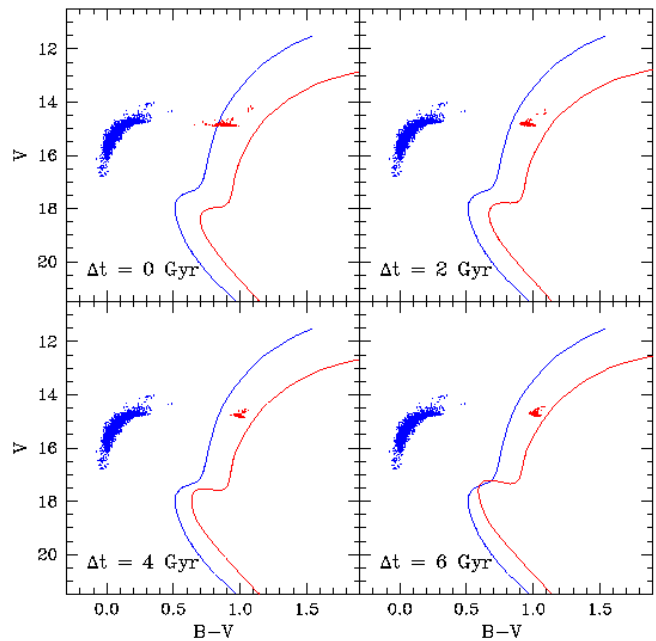}
\vspace{8.5cm}
\caption{Variation of HB morphology and MS turnoff of the most
metal-rich population (Z = 0.006) with respect to the most metal-poor population
(Z = 0.0006) under different assumptions regarding their relative ages}
\end{figure*}

This conclusion is in qualitative agreement with the age-metallicity relations
obtained directly from the MS turnoff stars (Hughes \& Wallerstein 2000;
Hilker \& Richtler 2000), indicating that $\omega$~Cen was self-enriched.
The strong resemblance of the populations in $\omega$~Cen to the Sagittarius
dwarf system, which is in the merging process with the Milky Way
(Ibata, Gilmore, \& Irwin 1994), then argues that $\omega$~Cen is nothing but
the relict of disrupted dwarf galaxy (Lee et al. 1999).

\begin{figure*}
\includegraphics{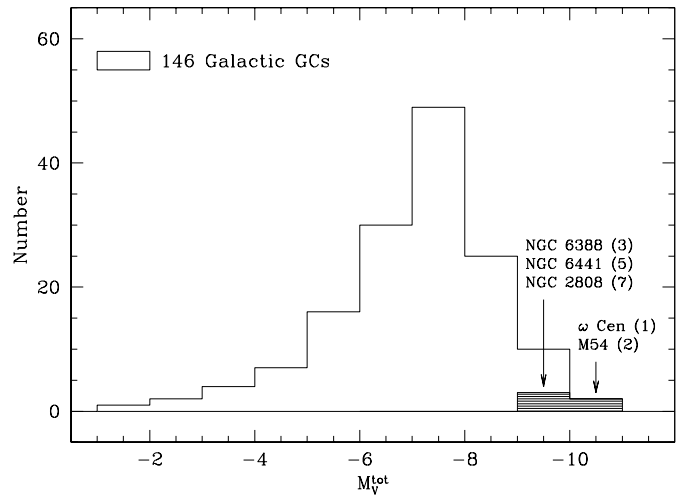}
\vspace{6.5cm}
\caption{Distribution of total brightness for 146 Galactic GCs from 
Harris (1996) catalog. Note that the clusters with bimodal HBs, NGC~6388,
NGC~6441, and NGC~2808, as well as $\omega$~Cen and M54 are among the most
massive clusters. The numbers in the parentheses denote their ranks in brightness.}
\end{figure*}

\section{Other Massive Globular Clusters: NGC~6388 and NGC~6441}

Interestingly enough, the clusters with bimodal HB, such as NGC~6388, NGC~6441,
and NGC~2808 are among the most bright (massive) ones next to $\omega$~Cen and M54
in their ranks (Figure 3).
In particular, NGC~6388 and NGC~6441 show color spreads among RGB stars which
appear to be larger than the effects expected from photometric errors and
differential reddenings.
We constructed population models in order to examine whether those features
 and the bimodal HBs could be induced by multiple populations.

Figure 4 presents the comparison of observed CMDs to our synthetic models for 
NGC~6388 and NGC~6441.
The observational error simulations included a small amount of differential
reddening (${\sigma}_{(B-V)}$ = 0.02), following the suggestions of 
Heitsch \& Richtler (1999) and Layden et al. (1999).
Our population models suggest the RGBs of these two clusters are broad enough
to contain two distinct populations within the observational uncertainties,
and the observed bimodal HBs are also readily explained with
the sum of blue HB from (relatively) metal-poor and older component and red HB
from metal-rich and younger component.
The age and metallicity spreads, required to reproduce their CMD morphologies,
are only about 1.2 Gyr and 0.15 dex for both clusters.

\begin{figure*}
\includegraphics{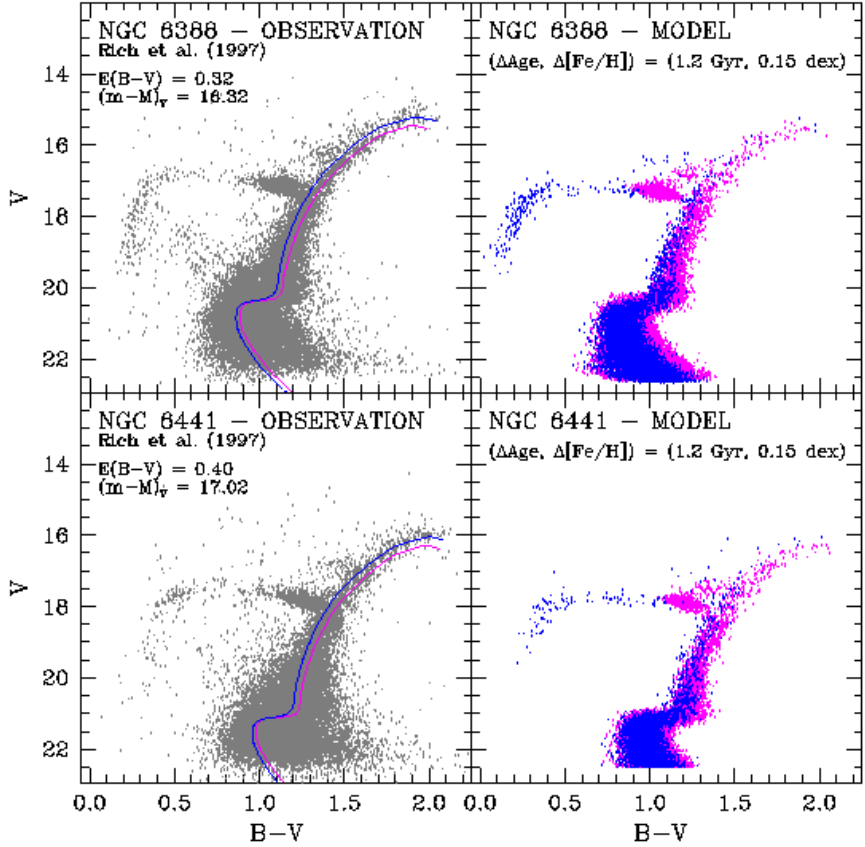}
\vspace{11.0cm}
\caption{Comparison of observations ($left$) and models ($right$) for NGC~6388 and
NGC~6441. }
\end{figure*}

\begin{figure*}
\includegraphics{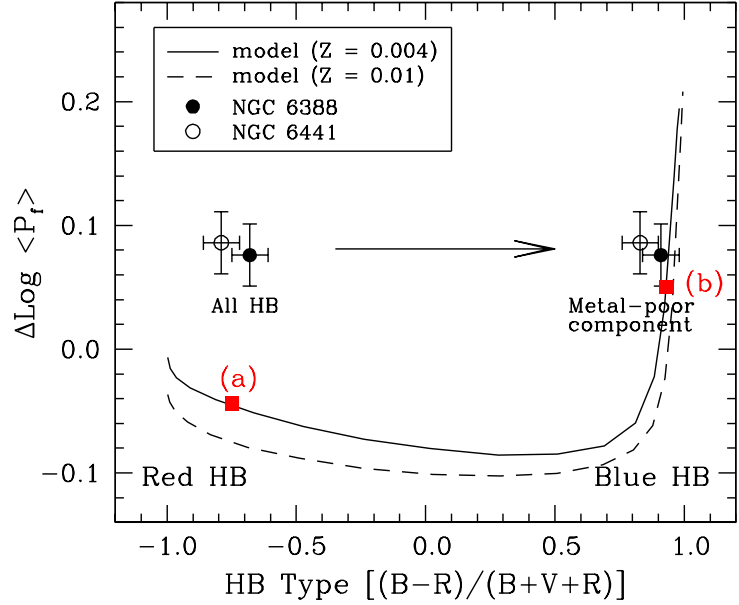}
\vspace{8.0 cm}
\caption{The mean periods of RR Lyraes in NGC~6388 and NGC~6441 are too
long for their metallicities. The models (solid \& dashed lines) provide a 
reasonable match, only if all RR Lyraes belong to metal-poor component, as 
predicted from our population models.}
\end{figure*}

\begin{figure*}
\includegraphics{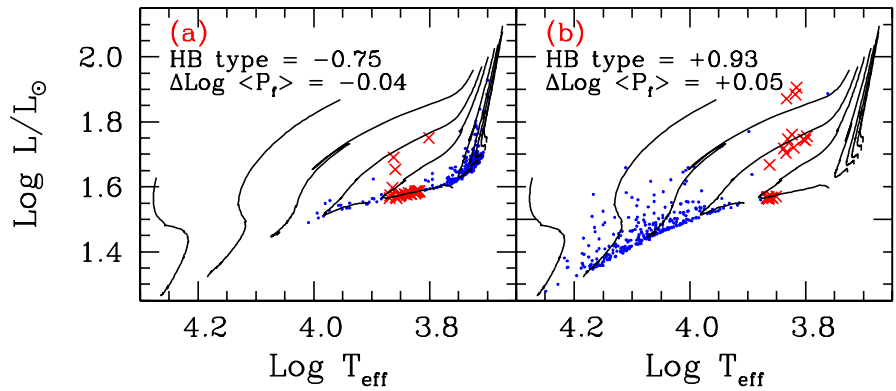}
\vspace{6.5cm}
\caption{HB models at two different HB types ($Z$ = 0.004) illustrate the evolutionary
states of RR Lyrae stars (crosses). The observed long periods (brighter luminosities)
are due to the evolutionary effect from the blue HB (panel b).}
\end{figure*}

The unusual properties of RR Lyraes in NGC~6388 and NGC~6441 also support this
view.
Recent observations (Pritzl et al. 2000) indicate that
(1) the fractions of c-type variables for these clusters have the 
characteristics of Oosterhoff II clusters, and 
(2) the mean pulsation periods of their ab-type variables are too long for 
their metallicities.

Figure 5 shows the variation of the mean fundamentalized period of RR Lyraes
with HB type as predicted from our synthetic HB models (see Demarque et al. 2000).
If all the HB stars are counted in, the HB types for NGC~6388 and NGC~6441 are
calculated to be $\sim$ -0.7 because of the dominent red HB population.
In this case, models can not explain the observed long periods because most 
RR Lyraes are near the zero-age HB (ZAHB) (see Figure 6a).
However, if all the RR Lyraes belong to the metal-poor component, as predicted
by our population models, the HB types become much bluer (close to +1).
The RR Lyraes in this case would be highly evolved stars from the blue HB,
and therefore much brighter than the stars near ZAHB (cf. Lee et al. 1990).
The longer periods of RR Lyraes are then naturally explained as a result of
higher luminosity (Figure 6b).
We believe this is a good evidence for two populations in these clusters.
More detailed investigations are in progress to see whether these two massive
clusters are in fact the relicts of disrupted dwarf galaxies, similar to the 
cases of $\omega$~Cen and M54.

\end{document}